\documentclass[11pt,fleqn,twocolumn,a4paper]{article}

\begin{document}

\noindent \begin{minipage}{16cm}{\noindent \Large Exact solutions of the Dirac equation and
induced representations   of the Poincar\'e group on the lattice}

\vspace{5mm}
{\noindent Miguel Lorente}

\vspace{5mm}
{\noindent Departamento de F\'{\i}sica, Universidad de Oviedo, 33007 Oviedo, Spain}

\vspace{5mm}
{\small We deduce the structure of the Dirac field on the lattice from the discrete version of differential
geometry and from the representation of the integral Lorentz transformations. The analysis of the
induced representations of the Poincar\'e group on the lattice reveals that they are reducible, a
result that can be considered a group theoretical approach to the problem of fermion doubling.}
\end{minipage}

\vspace{5mm}

\begin{list}{}{\setlength{\leftmargin}{5mm}}
\item [\bf 1.] {\bf DISCRETE DIFFERENTIAL \newline GEOMETRY}
\end{list}
\vspace{1mm}
Given a function of one independent variable the forward and backward differences are defined as
\begin{eqnarray*}
 \Delta f(x) &= &f(x + \Delta x) - f(x) \\
\nabla f(x) &=& f(x) - f(x - \Delta x)
\end{eqnarray*}
Similarly we define the average operators
\begin{eqnarray*}
\widetilde{\Delta }f(x) &= &{1 \over 2}\left\{{f(x+\Delta x) + f(x)}\right\} \\
\widetilde{\nabla }f(x) &=& {1 \over 2}\left\{{f(x)+ f(x-\Delta x)}\right\}
\end{eqnarray*}
This calculus can be enlarged to functions of several variables. In particular, we have for the
total difference operator
\begin{eqnarray*}
\lefteqn{\Delta f\left({x,y}\right)f\left({x+\Delta x,y+\Delta 
y}\right)-f\left({x,y}\right) =}\\
& &{\Delta }_{x}{\widetilde{\Delta }}_{y}f\left({x,y}\right)+{\widetilde{\Delta }}_{x}{\Delta
}_{y}f\left({x,y}\right)
\end{eqnarray*}
in an obvious notation. The finite increments of the variables $\Delta x,\Delta y \cdots $, can be used to
introduce a discrete differential form
\[\omega =a\left({x,y}\right)\Delta x+b\left({x,y}\right)\Delta y\]

\vspace*{60mm}

and the usual exterior product, from which we construct the p-form and the dual $(n-p)$ -form in the
n-dimensional space.

\begin{eqnarray*}
\sigma &=&{\frac{1}{p!}}{\sigma }_{{i}_{1}\cdots {i}_{p}}\Delta {x}^{{i}_{1}}\wedge \cdots \wedge \Delta
{x}^{{i}_{p}} \\ {\left({{}^*\sigma }\right)}_{{k}_{1}\cdots {k}_{n-p}}&=&{\frac{1}{p!}}{\sigma }^{{i}_{1}\cdots
{i}_{p}}{\varepsilon }_{{i}_{1}\cdots {i}_{p}{k}_{1}\cdots {k}_{p}}
\end{eqnarray*}

From the p-forms we can construct $(p+1)$-forms with the help of the exterior difference, for
instance,

\begin{eqnarray*}
\lefteqn{\Delta \omega \equiv \Delta a\wedge \Delta x+\Delta b\wedge \Delta y=} \\
& &\left({{\frac{\Delta_ {x}\tilde{\Delta }_{y}b}{\Delta x}}-{\frac{\tilde{\Delta }_x\Delta_y a}{\Delta
y}}}\right)\Delta x\wedge
\Delta y
\end{eqnarray*}

The exterior calculus can be used to express the laws of physics in terms of difference calculus,
such as classical electrodynamics, quantum field theory [1]. In particular, for the wave equation we
have $-{}^*\Delta {}^*\Delta \phi =0$, or

\begin{eqnarray*}
\lefteqn{\left\{-{\tilde\nabla }_{x}{\tilde{\nabla }}_{y}{\tilde{\nabla }}_{z}{\nabla }_{t}\tilde{\Delta }_{x}{\tilde{\Delta
}}_{y}{\tilde{\Delta }}_{z}{\Delta }_{t} \right. + }\\ 
&+&\hspace{-3mm}\left. {\nabla }_{x}{\tilde{\nabla }}_{y}{\tilde{\nabla }}_{z}{\tilde{\nabla }}_{t}{\Delta }_{x}{\tilde{\Delta
}}_{y}{\tilde{\Delta }}_{z}{\tilde{\Delta }}_{t}+\cdots \right\}\\
&.&\phi \left({xyzt}\right)=0
\end{eqnarray*}

\begin{list}{}{\setlength{\leftmargin}{5mm}}
\item [\bf 2.] {\bf EXACT SOLUTIONS FOR THE \newline DIRAC-EQUATION}
\end{list}
\vspace{1mm}

Given a scalar function $\Phi \left({{n}_{\mu }}\right)\equiv \left({{n}_{\mu }{\varepsilon }_{\mu }}\right)$
defined in the grid points of a Minkowski lattice with
elementary lengths ${\varepsilon }_{\mu }$, and the difference operators

\[{\delta }_{\mu }^{+}\equiv {\frac{1}{{\varepsilon }_{\mu }}}{\Delta }_{\mu }\prod\limits_{\nu \ \ne \ \mu }^{} {\tilde{\Delta
}}_{\nu },\quad {\delta }_{\mu }^{-}\equiv {\frac{1}{{\varepsilon }_{\mu }}}{\nabla }_{\mu }\prod\limits_{\nu \ \ne \ \mu
}^{} {\tilde{\nabla }}_{\nu }\]
\[{\eta }^{+}\equiv \prod\limits_{\mu }^{} {\tilde{\Delta }}_{\mu } ,\quad {\eta }^{-}\equiv \prod\limits_{\mu }^{}
{\tilde{\nabla }}_{\mu }, \quad \mu ,\nu =1,2,3,4\]
the Klein-Gordon equation can be read off

\begin{equation}
\left({{\delta }_{\mu }^{+}{\delta }^{{\mu }^{-}}-{m}_{0}^{2}{c}^{2}{\eta }^{+}{\eta }^{-}}\right)\Phi \left({{n}_{\mu
}}\right)=0
\end{equation}

Using the method of separation of variables, we obtain the exact solutions of this difference
equation

\begin{equation}
f\left({{n}_{\mu }}\right)=\prod\limits_{\mu \ =\ 0}^{3} {\left({{\frac{1-{\frac{1}{2}}i{\varepsilon }_{\mu }{k}_{\mu
}}{1+{\frac{1}{2}}i{\varepsilon }_{\mu }{k}_{\mu }}}}\right)}^{{n}_{\mu }}
\end{equation}
with ${k}_{\mu }$, continuous variables satisfiying the dispersion relations ${k}_{\mu }{k}^{\mu }={m}_{0}^{2}{c}^{2}$ 

The solutions (2) constitute a basis for the solutions of the Klein-Gordon equation.
If we impose spatial boundary conditions on the wave functions, we get

\[{k}_{i}={\frac{2}{{\varepsilon }_{i}}}\tan\ {\frac{\pi {m}_{i}}{N}},\ {m}_{i}=0,\cdots ,N-1\ ,\ i=1,2,3\]

With the help of this basis we can derive, as in the continuum case, Hamilton equations of motion, a
Fourier expansion for the Klein-Gordon fields [2]. We can apply the same technique to the Dirac
fields on the lattice. Starting from the Hamiltonian
\begin{eqnarray*}
\lefteqn{H={\varepsilon }_{1}{\varepsilon }_{2}{\varepsilon }_{3}\sum\limits_{{n}_{1}{n}_{2}{n}_{3}=0}^{N-1} {\tilde{\Delta
}}_{1}{\tilde{\Delta }}_{2}{\tilde{\Delta }}_{3}{\psi }^{+}(n) \ \times}\\
&\times &\left\{\gamma_0 {\gamma }_{1}{\frac{1}{{\varepsilon }_{1}}}{\Delta
}_{1}{\tilde{\Delta }}_{2}{\tilde{\Delta }}_{3}+\gamma_0 {\gamma }_{2}{\tilde{\Delta }}_{1}{\frac{1}{{\varepsilon }_{2}}}{\Delta
}_{2}{\tilde{\Delta }}_{3}\ +\right.\\
&+&\left.{\gamma }_{0}{\gamma }_{3}{\tilde{\Delta }}_{1}{\tilde{\Delta }}_{2}{\frac{1}{{\varepsilon
}_{3}}}{\Delta }_{3}+\gamma_0 {\tilde{\Delta }}_{1}{\tilde{\Delta }}_{2}{\tilde{\Delta }}_{3}\right\}{\psi }_{(n)}
\end{eqnarray*}
we obtain from the Hamilton equations of motion, the Dirac equation
\begin{equation}
\left({i{\gamma }_{\mu }{\delta }^{\mu +}-{m}_{0}c{\eta }^{+}}\right)\psi \left({{n}_{\mu }}\right)=0
\end{equation}
and from this we recover the Klein-Gordon equation. The transfer matrix, which carries the Dirac
field from one time to the next can be obtained from the evolution operator [3]
\[U={\frac{1+{\frac{1}{2}}i{\varepsilon }_{0}H}{1-{\frac{1}{2}}i{\varepsilon }_{0}H}}\ \ \ ,\ \ \ \psi
\left({{n}_{0}+1}\right)=U\psi \left({{n}_{0}}\right){U}^{\dag }\]

A basis for the solutions to the Dirac equation can be obtained from the functions (2) and the
standard four components spinors. Our model for the fermion fields satisfies the following
conditions [2]:
\begin{list}{}{\setlength{\parsep}{0mm}\setlength{\topsep}{0mm}\setlength{\leftmargin}{6.5mm}}
\item[i)] the Hamiltonian is translational invariant
\item[ii)] the Hamiltonian is hermitian
\item[iii)] for $m_0=0$, the wave equation is invariant under global chiral transformations
\item[iv)] there is no fermion doubling
\item[v)] the Hamiltonian is non-local \newline
\end{list}
Finally, coupling the vector field to the electro-magnetic vector potential we construct a gauge
invariant vector current leading to the correct axial anomaly [2]
\vspace{5mm}

\begin{list}{}{\setlength{\leftmargin}{5mm}}
\item [\bf 3.] {\bf INTEGRAL LORENTZ \newline TRANSFORMATIONS}
\end{list}

A group theoretical approach to the Dirac equation, can be given from the induced representations
of Poincar\'e group on the lattice [4]. We start with the integral transformations of the complete
Lorentz group generated by the Kac generators

\setlength{\arraycolsep}{1mm}
\setlength{\tabcolsep}{0.5mm}
{\noindent\begin{tabular}{ll}
${S}_{1}=\left(\begin{array}{rrrr} 1&0&0&0\\
0&0&1&0\\
0&1&0&0\\
0&0&0&1\end{array}\right)$,&${S}_{2}=\left(\begin{array}{rrrr}1&0&0&0\\
0&1&0&0\\
0&0&0&1\\
0&0&1&0\end{array}\right)$ \\ [8mm]
${S}_{3}=\left(\begin{array}{rrrr}1&0&0&0\\
0&1&0&0\\
0&0&1&0\\
0&0&0&-1\end{array}\right)$,&${S}_{4}=\left(\begin{array}{rrrr}2&1&1&1\\
-1&0&-1&-1\\
-1&-1&0&-1\\
-1&-1&-1&0\end{array}\right)$
\end{tabular}
}

Any integral matrix of the complete Lorentz group can be factorized

\[L={P}_{1}^{\eta }{P}_{2}^{\theta }{P}_{3}^{i}{S}_{4}\ldots {S}_{4}{P}_{1}^{\rm
\delta }{P}_{2}^{\varepsilon }{P}_{3}^{\zeta }{S}_{4}\left\{{{S}_{1}^{\alpha
}{S}_{2}^{\beta }{S}_{3}^{\gamma }}\right\}_{\rm perm.}\]
where ${P}_{1}={S}_{1}{S}_{2}{S}_{3}{S}_{2}{S}_{1}$,\quad ${P}_{2}={S}_{2}{S}_{3}{S}_{2}$,\newline
${P}_{3}={S}_{3}$,\qquad  $\alpha,\beta,\ldots =0,1$

The finite representations of the Lorentz group can be obtained by standard method. In
particular, for the two dimensional representation $\alpha \in SL(2,C)$ and the parity operator $I_s$ we obtain the
(reducible) Dirac representation [5]

\[D\left(\alpha ,{I}_{s}\right)=\left(\begin{array}{cc}\alpha &0\\
0&{\left({\alpha }^{\dag }\right)}^{-1}\end{array}\right)\otimes \left(\begin{array}{cc}0&1\\
1&0\end{array}\right)\]

In order to get $UIR$, we introduce the projection operator

\[Q\left({p}\right)={\frac{1}{2}}\left({I+W\left({p}\right)}\right)\; ,\; W\left({p}\right)\equiv
{\frac{1}{{m}_{0}c}}{\gamma }^{\mu }{p}_{\mu }\]

Applying this operator to a four component spinor we obtain the Dirac equation in momentum space

\begin{equation}
\left({{\gamma }^{\mu }{p}_{\mu }-{m}_{0}cI}\right)\psi \left({{p}_{\mu }}\right)=0
\end{equation}

We recover the Dirac equation in position space (3) by making the substitution $p_\mu={\frac{2}{\varepsilon }}\tan\ \pi
{k}_{\mu }{\varepsilon }_{\mu }$ in (4) and then applying the discrete Fourier transform

\begin{equation}
F(n)=\int_{-1/2\varepsilon }^{1/2\varepsilon }\exp\ \left({-2\pi ikn\varepsilon }\right)\hat{F}\left({k}\right)dk
\end{equation}
\vspace{3mm}

\begin{list}{}{\setlength{\leftmargin}{5mm}}
\item [\bf 4.] {\bf INDUCED REPRESENTATIONS OF THE POINCAR\'E GROUP}
\end{list}
\vspace{1mm}

Given an integral Lorentz transformation $\Lambda$ and discrete translation characterized by a
four-vector $a$, the induced representations are given by standard methods [5]

\begin{list}{}{\setlength{\parsep}{0mm}\setlength{\topsep}{0mm}}
\item[i)] choose an $UIR$, ${D}^{\stackrel{o}{k}}\left({a}\right)$ of the translation group (the kernel of the
discrete Fourier transform) (5))
\item[ii)] define the little group $H\in SO(3,1)$ by the stability condition: $D \left(ha\right)=D\left(a\right)  ,  h\in H$
\item[iii)] construct for the group $T_4\times_sH$ the $UIR$ \newline
${D}^{\stackrel{o}{k} ,\alpha }\left({a,h}\right)
={D}^{\stackrel{o}{k}}\left({a}\right)\otimes {D}^{\alpha }\left({h}\right)$
\item[iv)] choose coset generators $c,c'$ such that $\left({\tilde{a},\tilde{\wedge }}\right)c=c'\left({a,h}\right)$. For the coset
representatives we can choose $\wedge \equiv L\left({k}\right)$ that takes $\stackrel{o}{k}$ into an arbitrary integral vector of
the unit hyperboloid. Finally we have \newline
${D}^{\stackrel{o}{k},\alpha }_{k,k'}\left({\tilde{a},\tilde{\wedge
}}\right)={D}^{k'}\left({\tilde{a}}\right){D}^{\alpha }\left({{L}^{-1}\left({k'}\right)\tilde{\wedge }L\left({k}\right)}\right)$
\end{list}
\vspace{3mm}

\begin{list}{}{\setlength{\leftmargin}{5mm}}
\item [\bf 5.] {\bf IRREDUCIBILITY AND FERMION DOUBLING}
\end{list}
\vspace{1mm}

The dual group of the discrete translation group is given by all the points from the Brillouin zone.
We characterize the discrete orbit of the point group by functions on the dual space. We
formulate the following conditions for these constraints:
\begin{list}{}{\setlength{\parsep}{0mm}\setlength{\topsep}{0mm}}
\item[i)] they should vanish on and only on the points of the orbit
\item[ii)]they should admit a periodic extension
\item[iii)] they must be Lorentz invariant
\item[iv)]the difference equations should go to the continuous wave equation when the lattice spacing
goes to zero.\newline
\end{list}
The difference equation (3) satisfies iv) but leads to the constraints
\[{\frac{4}{{\varepsilon }^{2}}}\left(\tan\ \pi {k}_{\mu }\varepsilon \right) \left(\tan\ \pi {k}^{\mu }\varepsilon\right)
-{m}_{0}^{2}{c}^{2}=0\] that violate conditions i) and iii), therefore the representations characterized by these
constraints are reducible.

In the problem of fermion doubling one considers the inverse of the propagator, such as
\[{\frac{4}{{\varepsilon }^{2}}}\left({\sin\ \pi {k}_{\mu }\varepsilon }\right)\ \left({\sin\ \pi {k}_{\mu }\varepsilon
}\right)-{m}_{0}^{2}{c}^{2}\]
is invariant under the cubic group and become zero on the points of the orbit under this group,
but is zero also in the border of the Brillouin zone $-1/\varepsilon \le {k}_{\mu }\le 1/\varepsilon$. Therefore the
representation is not irreducible and can be reduced giving rise to different elementary systems with the same mass.

\vspace{9mm}
\newpage
{\noindent \bf REFERENCES}
\vspace{4mm}

\begin{list}{}{\setlength{\parsep}{0mm}\setlength{\topsep}{0mm}\setlength{\leftmargin}{5mm}}
\item [1.] M.Lorente in Symmetry Methods in Physics (N. Sissakian, G.S. Pogosyan ed.) Dubna, 1996, p.
368-77.
\item [2.] M. Lorente, J. Group Th. Phys. 1 (1993) 105-121.
\item [3.] M. Lorente, Lett. Math. Phys, 13 (1987) 229-236; Phys. Lett. B 232 (1989) 345-350.
\item [4.] M. Lorente in Symmetries in Science IX (B. Gruber, M. Ramek ed.) Plenum, N.Y. 1997, p. 211-223.
\item [5.] M. Lorente, P. Kramer in Symmetries in Science X (B. Gruber, M. Ramek ed) Plenum, N.Y. 1998 p. 179-195
	\end{list}

\end{document}